\documentclass[12pt]{article}

\usepackage[utf8]{inputenc}
\usepackage[american]{babel}

\usepackage[margin=1in]{geometry}
\usepackage{setspace}
\usepackage{titlesec}
\titlespacing*{\section}{0pt}{0pt}{0pt}
\titlespacing*{\subsection}{0pt}{0pt}{0pt}
\titlespacing*{\subsubsection}{0pt}{0pt}{0pt}

\titleformat{\section}[block]{\normalfont\large\bfseries}{\thesection}{1em}{}

\titleformat{\subsection}[block]{\normalfont\normalsize\bfseries}{\thesubsection}{1em}{}

\titleformat{\subsubsection}[block]{\normalfont\small\itshape}{\thesubsubsection}{1em}{}

\usepackage[most]{tcolorbox}

\usepackage{natbib}
\usepackage{csquotes}

\title{Recommender systems, stigmergy, and the tyranny of popularity} 

\author{
  Zackary Okun Dunivin\textsuperscript{1}\thanks{Corresponding author: zdunivin@ucdavis.edu}\\
  \vspace{.5em}
  Paul Smaldino \textsuperscript{2,3}\\
  \textsuperscript{1}University of California, Davis\\
  \textsuperscript{2}University of California, Merced\\
  \textsuperscript{3}Santa Fe Institute\\
}

\date{\today} 

\begin{document}
\maketitle

\setlength{\parskip}{1em}

\begin{abstract}
Scientific recommender systems, such as Google Scholar and Web of Science, are essential tools for discovery. Search algorithms that power work through stigmergy, a collective intelligence mechanism that surfaces useful paths through repeated engagement. While generally effective, this ``rich-get-richer'' dynamic results in a small number of high-profile papers that dominate visibility. This essay argues argue that these algorithm's over-reliance on popularity fosters intellectual homogeneity and exacerbates structural inequities, stifling innovative and diverse perspectives critical for scientific progress. We propose an overhaul of search platforms to incorporate user-specific calibration, allowing researchers to manually adjust the weights of factors like popularity, recency, and relevance. We also advise platform developers on how text embeddings and LLMs could be implemented in ways that increase user autonomy. While our suggestions are particularly pertinent to aligning recommender systems with scientific values, these ideas are broadly applicable to information access systems in general. Designing platforms that increase user autonomy is an important step toward more robust and dynamic information ecosystems.

\end{abstract}

\medskip
\noindent
\textbf{Keywords:} search algorithms; information retrieval; human-computer interaction; artificial intelligence; complex systems

\newpage
\section*{Introduction}
Success in both cultural and intellectual markets is not driven by quality alone. Instead, visibility and accessibility increase salience and lower search costs, boosting the popularity of high-profile items and creating a cycle in which popularity begets more popularity. Take the example of the Music Lab experiment \citep{salganik2006inequality}, which constructed a simplified recommender system for songs, and found that songs that were initially advertised on the landing page were far more likely to be played, largely independent of quality. This further increased the visibility of those songs, and eventually led to the sort of massive inequality in exposure to which we've become increasingly accustomed (Frank 2011). This experiment highlights how recommender systems can shape patterns of consumption, consequently influencing tastes, affordances, and the surrounding culture.

Recommender systems have become a ubiquitous element of daily life, governing access to media, consumer goods, and information more generally. Over time, the content of our personal and professional lives has become increasingly shaped by the algorithms at their core. These systems are also key drivers of scientific research, shaping how we discover existing ideas, methods, and findings. Scientific recommender systems such as Google Scholar and Web of Science don't only facilitate scientific research; they shape its content. It is therefore worth reflecting on the role of these systems in scientific knowledge production, assessing the implicit values they reflect, and considering how we might alter them to better support good science.

\section*{Stigmergy Drives Recommender Systems}
Recommender systems work because of stigmergy, a process of indirect coordination in human and non-human social systems \citep{theraulaz1999stigmergy}. Ant pheromone trails, for instance, exemplify how local interactions produce emergent patterns of collective behavior without centralized control. As each ant follows the strongest pheromone gradient it can sense while laying down more pheromones of its own, the colony's movement patterns become increasingly aligned until all the ants eventually travel along the same narrow path.

In recommender systems like Google Search and Google Scholar, stigmergy manifests as preferential attachment \citep{parunak2005survey}. Highly cited or frequently accessed items are disproportionately surfaced, creating heavy-tailed distributions where a small number of items dominate attention---the rich get richer \citep{frieze2006search}. Such distributions can have advantages. They may reflect genuine differences in quality or utility, helping shared vocabularies and paradigms coalesce across communities. These collective frameworks enable large-scale scientific coordination by ensuring researchers encounter a common set of references.

However, the same dynamics can also exacerbate inequality and homogeneity. The extreme ``heaviness'' of the tail in the distribution of influence---the degree of concentration of resources at the top---raises concerns. Overreliance on recommender systems narrows the field of visibility, crowding out less mainstream ideas. In science, this reduces access to diverse perspectives, methods, and paradigms, which in turn stifles innovation \citep{foster2015tradition,smaldino2022interdisciplinarity}.

\section*{The Costs of Inequality and Homogeneity in Science}
Heavy-tailed distributions of influence in science contribute to entrenched hierarchies. Articles with high citation counts reinforce the reputations of their authors, enabling them to attract more attention and funding \citep{bol2018matthew}. This cycle also favors dominant paradigms, making it difficult for alternative perspectives to gain traction.

The implications are broad. Equity and diversity are essential for social and scientific progress. Diversity enables the exploration of a larger solution space, improving collective problem-solving and creating opportunities to synthesize ideas across disciplines \citep{smaldino2024transient}. It also mitigates structural inequities that disproportionately affect underrepresented groups. Scholars from marginalized backgrounds face greater barriers to drawing on professional networks that catalyze careers \citep{forret2006impact,pedulla2019race}; a study by \cite{hofstra2020diversity} found that minority scholars' contributions were less likely to find academic positions, despite, and in part due producing more innovative research. While recommender systems may seem like an unbiased and preferable alternative to ``old boy networks,'' in truth networks and recommender systems work in concert to reproduce inequity. Without early visibility, structurally marginalized scholars' contributions may be less impactful and unrewarded in faculty placement, advancement, and grant-funding, perpetuating systemic injustice.

Diversity is not without trade-offs. Shared language and aligned goals facilitate collaboration and productivity. And a subordinate or peripheral network of innovators may be critical for maintaining the diversity that disproportionately benefits the mainstream \citep{milzman2023core,turner2023minority}---diversity does not necessarily imply equity. Yet, overemphasis on similarity narrows the landscape of ideas. Excessive inequality and lack of diversity harm both individuals and society, as innovative contributions from less prominent voices may be stifled. A healthier ecosystem therefore requires tempering, not flattening, inequality across levels of scientific organization, from single articles to entire disciplines.

\section*{Rethinking Recommender Systems}
What can we do to counteract the tyranny of popularity? Researchers could deliberately engage with less cited papers that articulate similar ideas with comparable cogency, and include less prominent researchers in collective efforts. However, individual actions alone are unlikely to shift systemic patterns without coordination. System-level changes are needed, and they are feasible through an overhaul of search algorithms and their user-interfaces.

Recommender systems, being proprietary, generally obscure their inner workings. Nevertheless, certain mechanisms and biases are evident, and these can guide our interventions. At the same time, any proposal must consider the financial and structural incentives that constrain the corporations operating these platforms.

In commercial contexts, recommender systems are calibrated to maximize engagement. For instance, music streaming platforms promote certain content---sometimes paid for by producers---while aiming to keep users listening. This dynamic has roots in early commercial radio's ``payola'' practices. Similarly, general search engines like Google balance relevance with monetization, incorporating click-through rates and ``sponsored content''---paid advertisements---into their ranking systems. 

By contrast, research platforms such as Google Scholar or Web of Science, are less likely to employ direct monetization, but biases remain. These systems may, for instance, over-prioritize highly cited papers, as doing so may reduce user complaints by surfacing ``safe,'' broadly accepted results despite the hidden cost of missing more apt or innovative works along with some obvious duds. This can lead to the over-association of particular keywords with a narrow set of papers, leaving other relevant but less cited works and approaches underexplored. More importantly, the algorithm's prioritization of highly cited results intensifies stigmergy, in which the rich continue to get richer.

The search algorithms underlying these systems balance various features, including query relevance, popularity, recency, and user's content preferences. Often these features are in tension, with rankings reflecting trade-offs intended to maximize user satisfaction or engagement. We can imagine all of these factors blended into a single metric, reflecting the weight of the algorithm's inferred value of a result. In this case, downweighting popularity would promote diversity system-wide. Another approach is to add stochastic noise to the popularity score of each item, disrupting deterministic feedback loops and diversifying the outcomes presented to users. Indeed, adding noise to all factors---perceived relevance, recency, and user preferences---is likely to both increase diversity and, most importantly, offer results that yield genuine insight.  While a highly skewed distribution of influence is probably inevitable \citep{price1976general} and perhaps even desirable to some extent, noisy interventions can work to severely reduce the steepness of that distribution. Alternatively, users might select an algorithm that explicitly biases sampling toward middling-popularity items, avoiding the steep heavy-tailed dynamics that reinforce entrenched hierarchies. The key feature here is to provide user control over the algorithm's emphasis on popularity and related factors. Sometimes we need to find the most commonly cited source on a topic, while other times it benefits us to prioritize exploration over exploitation.

\section*{Benefits of User-Specific Calibration}
Instead of relying solely on fixed algorithms, platforms should introduce user-specific calibration tools, empowering individuals to adjust the relative weights of variables like popularity, recency, and relevance. For example, researchers seeking obscure but relevant articles could downweight popularity, while those exploring emerging trends might prioritize recency. Adjustments could be made at the level of individual searches, enabling users to refine overly generic results or tailor recommenders to reflect specific disciplinary needs.

Such customization could also be automated. For users uninterested in manual adjustments, systems could iteratively refine recommendations by sweeping algorithmic parameters and soliciting feedback. For example, a search engine might experiment with varying the weight of popularity or recency and ask users to evaluate the quality of the results. This process would align the system with an individual's preferences over time. Such automated customization would be easily accomplished using widespread machine learning tools. 

Opening search algorithms to user-specific calibration has the potential to disrupt the stigmergic feedback loops that currently reinforce inequality and homogeneity. By diversifying outcomes, these systems could better serve niche user needs while maintaining engagement. Importantly, this would not preclude platforms from pursuing commercial goals, as adjustments could be confined to a subset of ``power users'' who are willing to spend extra time calibrating their search. While most users probably won't engage deeply with calibration tools, researchers would likely benefit disproportionately, given their highly specialized needs, engagement in critical thought, and reliance on search systems to achieve professional goals.

Moreover, user-specific calibration addresses a fundamental flaw in current systems: their one-size-fits-all approach. By allowing adjustments at the individual level, platforms could accommodate a wider range of preferences and objectives, catering to a subset of the market that is currently underserved. In other words, platforms would ultimately benefit as well. 

\section*{Search in the Time of LLMs}
Algorithmic discovery is in the midst of its most dramatic transformation since PageRank, driven by large language models (LLMs) \citep{metzler2021rethinking}. New layers, such as Perplexity.ai, the ``Deep Research'' features of LLM platforms, and AI-overviews of search results process the outputs of traditional indexes, filtering, summarizing, and sometimes obscuring the route from query to source. Shah and Bender's \citeyearpar{shah2024envisioning} review of AI-driven information access systems warns that these layers can homogenize results and dull users' critical edge by serving pre-chewed summaries while the underlying literature recedes from view. They also proffer principles---link persistence, plurality, transparency---that a healthy systems must satisfy. 
We endorse these principles in the following suggestions detailing potential roles for language models in scientific search and search more broadly. We reject language models as oracles that replace reading; instead we cautiously suggest enlisting them as linguistic interfaces that translate sophisticated research intentions into richer retrieval mechanisms.

\subsection*{First, semantic retrieval beyond keywords}
Transformer-powered document embeddings capture conceptual similarity that keyword match misses. These embeddings let a query on ``stigmergic feedback in scholarly communication'' surface adjacent work in evolutionary biology or computer-supported cooperative work—papers whose titles may share no tokens with the query. Empirical benchmarks show that such semantic retrieval widens the candidate set without sacrificing precision \cite{metzler2021rethinking}. This enlarges the exploration frontier by reducing reliance on direct keyword matches, instead capturing the gestalt of a query.

\subsection*{Second, LLM-assisted search parameter control}
While explicit sliders may feel unnatural, a researcher could direct an LLM agent to adjust the parameter weights according to their preferences. Multi-sentence natural language queries could include both topical directives and requests for "less popular" or ``very recent'' publications. Although mediated by an LLM, the search engine would return direct links to indexed documents, so provenance is accessible and audit-ready, and hallucinations should be minimal or non-existent.

\subsection*{Third, interactive orchestration with an AI assistant}
Consider a brief exchange:

\begin{tcolorbox}[
  colback=white,
  colframe=black!50,
  boxrule=0.5pt,
  arc=2pt,
  left=1em,
  right=1em,
  top=1em,
  bottom=1em
]
\textbf{Researcher:} I'm modeling ``inadvertent coordination'' in open‐source communities. Show empirical studies using theory from organizational ecology or animal collective behavior, but avoid works by [top‐cited three authors in that niche].\\

\textbf{LLM interface:} \textit{*performs initial semantic search*} After analyzing the top 100 results (links provided), I see that ecology dominates, but there are also studies in animal behavior. I also found several ethnographic studies of craft guilds---shall I emphasize the anthropological scope to boost diversity? I'm noticing that there is a proliferation of studies after 2011, perhaps sparked by [prominent 2009 publication].\\

\textbf{Researcher:} Yes, expand our search to include work related to craft guilds and other literatures that are linked to this work. No need to adjust the effect of recency, I expect that these papers will surface naturally. But let's increase the priority of low‐citation papers when they come from outside ecology or animal behavior.
\end{tcolorbox}

\noindent The model's role is to characterize search results, and suggest parameter adjustments and frames for semantic search, then wait for human judgment. Topical fit is still ultimately determined by the scholar reading abstracts and articles; the LLM merely accelerates iteration.

These capabilities come with increased costs. Training and running LLMs consume significant energy and rely on labor-intensive data pipelines. We suggest that platforms adopting approaches advocated here contextualize search platform options with carbon footprint and data-provenance statements. Yet, when provenance, plurality, and agency are privileged, LLMs can become instruments to increase diversity and improve search specificity. Semantic reach finds the long-tail papers; dialogic refinement lets scholars steer the search; database-backed links guarantee auditability. These elements translate our general call for user-directed diversity into a concrete workflow that enlarges the horizon of scientific discovery while mitigating legitimate concerns of the harms of AI-driven information systems. Whether large AI models like LLMs represent a net benefit or net cost to humanity is open to debate. But if we are to live in a world with LLMs, let us use them to benefit science and humanity as best as we are able.

\section*{Better Science through Better Search}

Scientists strive to produce knowledge that is rigorous, useful, and effectively communicated. Recommender systems should align with these goals, balancing the need for diversity with the pragmatic realities of coordination and the functioning aspects of meritocracy. It is well established in evolutionary theory that the pace of adaptation is proportional to the variance within a population \citep{fisher1930genetical}. Similarly, maintaining intellectual diversity hedges against future challenges, ensuring resilience and innovation. Recalibrating search to dampen stigmergy and expanding user control over search, with that control enhanced by LLMs, will increase the diversity of results and the opportunity for innovation. Our suggestion represents an opportunity to improve both the individual experience of doing science and the collective endeavor of scientific discovery in one fell swoop.

\bibliography{ref}

\begin{thebibliography}{}

\bibitem[Bol et~al., 2018]{bol2018matthew}
Bol, T., {de Vaan}, M., and {van de Rijt}, A. (2018).
\newblock The {Matthew} effect in science funding.
\newblock {\em Proceedings of the National Academy of Sciences}, 115(19):4887--4890.

\bibitem[{de Solla Price}, 1976]{price1976general}
{de Solla Price}, D. (1976).
\newblock A general theory of bibliometric and other cumulative advantage processes.
\newblock {\em Journal of the American Society for Information Science}, 27(5):292--306.

\bibitem[Fisher, 1930]{fisher1930genetical}
Fisher, R.~A. (1930).
\newblock {\em The Genetical Theory of Natural Selection}.
\newblock Clarendon Press.

\bibitem[Forret, 2006]{forret2006impact}
Forret, M.~L. (2006).
\newblock The impact of social networks on the advancement of women and racial/ethnic minority groups.
\newblock {\em Gender, Ethnicity, and Race in the Workplace}, 3:149--166.

\bibitem[Foster et~al., 2015]{foster2015tradition}
Foster, J.~G., Rzhetsky, A., and Evans, J.~A. (2015).
\newblock Tradition and innovation in scientists’ research strategies.
\newblock {\em American Sociological Review}, 80(5):875--908.

\bibitem[Frieze et~al., 2006]{frieze2006search}
Frieze, A., Vera, J., and Chakrabarti, S. (2006).
\newblock The influence of search engines on preferential attachment.
\newblock {\em Internet Mathematics}, 3(3):361--381.

\bibitem[Hofstra et~al., 2020]{hofstra2020diversity}
Hofstra, B., Kulkarni, V.~V., Munoz-Najar~Galvez, S., He, B., Jurafsky, D., and McFarland, D.~A. (2020).
\newblock The diversity–innovation paradox in science.
\newblock {\em Proceedings of the National Academy of Sciences}, 117(17):9284--9291.

\bibitem[Metzler et~al., 2021]{metzler2021rethinking}
Metzler, D., Tay, Y., Bahri, D., and Najork, M. (2021).
\newblock Rethinking search: {Making} domain experts out of dilettantes.
\newblock {\em ACM Special Interest Group on Information Retrieval Forum}, 55(1):1--27.

\bibitem[Milzman and Moser, 2023]{milzman2023core}
Milzman, J. and Moser, C. (2023).
\newblock Decentralized core-periphery structure in social networks accelerates cultural innovation in agent-based model.
\newblock In {\em Proceedings of the 22nd International Conference on Autonomous Agents and Multiagent Systems (AAMAS2023)}, London, United Kingdom.
\newblock May 29–June 2 2023, IFAAMAS.

\bibitem[Pedulla and Pager, 2019]{pedulla2019race}
Pedulla, D.~S. and Pager, D. (2019).
\newblock Race and networks in the job search process.
\newblock {\em American Sociological Review}, 84(6):983--1012.

\bibitem[Salganik et~al., 2006]{salganik2006inequality}
Salganik, M.~J., Dodds, P.~S., and Watts, D.~J. (2006).
\newblock Experimental study of inequality and unpredictability in an artificial cultural market.
\newblock {\em Science}, 311(5762):854--856.

\bibitem[Shah and Bender, 2024]{shah2024envisioning}
Shah, C. and Bender, E.~M. (2024).
\newblock Envisioning information access systems: {What} makes for good tools and a healthy {Web?}
\newblock {\em ACM Transactions on the Web}, 18(3):1--24.

\bibitem[Smaldino et~al., 2024]{smaldino2024transient}
Smaldino, P.~E., Moser, C., Pérez~Velilla, A., and Werling, M. (2024).
\newblock Maintaining transient diversity is a general principle for improving collective problem solving.
\newblock {\em Perspectives on Psychological Science}, 19(2):454--464.

\bibitem[Smaldino and O'Connor, 2022]{smaldino2022interdisciplinarity}
Smaldino, P.~E. and O'Connor, C. (2022).
\newblock Interdisciplinarity can aid the spread of better methods between scientific communities.
\newblock {\em Collective Intelligence}, 1(2):26339137221131816.

\bibitem[Theraulaz and Bonabeau, 1999]{theraulaz1999stigmergy}
Theraulaz, G. and Bonabeau, E. (1999).
\newblock A brief history of stigmergy.
\newblock {\em Artificial Life}, 5(2):97--116.

\bibitem[Turner et~al., 2023]{turner2023minority}
Turner, M.~A., Singleton, A.~L., Harris, M.~J., Harryman, I., Lopez, C.~A., Arthur, R.~F., Muraida, C., and Jones, J.~H. (2023).
\newblock Minority-group incubators and majority-group reservoirs support the diffusion of climate change adaptations.
\newblock {\em Philosophical Transactions of the Royal Society B}, 378(1889):20220401.

\bibitem[{van Dyke Parunak}, 2005]{parunak2005survey}
{van Dyke Parunak}, H. (2005).
\newblock A survey of environments and mechanisms for human-human stigmergy.
\newblock In {\em Proceedings of the International Workshop on Environments for Multi-Agent Systems}, pages 163--186.

\end{thebibliography}

\end{document}